\documentclass[10pt]{article}
\usepackage{graphicx}
\usepackage{amsmath}
\usepackage{amssymb}
\usepackage{caption2}
\begin{document}
\begin{center}
{\large {\bf \sc{  Axialvector tetraquark candidates for the $Z_c(3900)$, $Z_c(4020)$, $Z_c(4430)$, $Z_c(4600)$ }}} \\[2mm]
Zhi-Gang  Wang \footnote{E-mail: zgwang@aliyun.com.  }    \\
 Department of Physics, North China Electric Power University, Baoding 071003, P. R. China
\end{center}

\begin{abstract}
In this paper, we construct the axialvector and tensor current operators to investigate   the ground state tetraquark states and the first radially excited tetraquark states  with the quantum numbers $J^{PC}=1^{+-}$ via the QCD sum rules systematically, and observe that there are one axialvector tetraquark candidate for the $Z_c(3900)$ and $Z_c(4430)$, two axialvector tetraquark candidates for the $Z_c(4020)$,  three axialvector tetraquark candidates  for the $Z_c(4600)$.
\end{abstract}



\section{Introduction}

In 2019, the LHCb collaboration performed an angular analysis of the weak decays $B^0\to J/\psi K^+\pi^-$ using proton-proton collision data, examined   the $m(J/\psi \pi^-)$ versus the $m(K^+\pi^-)$ plane, and observed two possible resonant structures in the vicinity of  the energies $m(J/\psi \pi^-)=4200 \,\rm{MeV}$ and $4600\,\rm{MeV}$, respectively \cite{LHCb-Z4600}.
  There have been two tentative assignments of the  structure $Z_c(4600)$ in the vicinity of $m(J/\psi \pi^-)=4600 \,\rm{MeV}$, the
  $[dc]_P[\bar{u}\bar{c}]_A-[dc]_A[\bar{u}\bar{c}]_P$ type vector tetraquark state with $J^{PC}=1^{--}$ \cite{Wang-Z4600-V} and the first radially excited  $[dc]_T[\bar{u}\bar{c}]_A-[dc]_A[\bar{u}\bar{c}]_T$ type  tetraquark state with $J^{PC}=1^{+-}$  \cite{ChenHX-Z4600-A}. In this paper, we use the subscripts $P$, $S$,  $V$, $A$ and $T$ to represent  the pseudoscalar, scalar,  vector, axialvector and tensor color-antitriplet  diquark states, respectively.

 In 2013, the BESIII collaboration  observed the charged charmonium-like  resonance  $Z^{\pm}_c(3900)$ in the $\pi^\pm J/\psi$ invariant mass spectrum in the process $e^+e^- \to J/\psi \pi^+\pi^-$ with $M_{Z_c}=(3899.0\pm 3.6\pm 4.9)\,\rm{ MeV}$
and  $\Gamma_{Z_c}=(46\pm 10\pm 20) \,\rm{MeV}$, respectively \cite{BES3900}. The Belle collaboration also observed the $Z^\pm_c(3900)$ in the same process   \cite{Belle3900}, furthermore,   the  CLEO collaboration  confirmed the existence of the $Z_c^\pm(3900)$  \cite{CLEO3900}.
 Almost at the same time, the BESIII collaboration  observed
the charmonium-like resonance $Z^{\pm}_c(4025)$ in the vicinity of the  threshold $(D^{*} \bar{D}^{*})^{\pm}$  in the electron-positron
scattering process $e^+e^- \to (D^{*} \bar{D}^{*})^{\pm} \pi^\mp$ \cite{BES1308}. Moreover, the  BESIII collaboration observed the charmonium-like resonance $Z_c^\pm(4020)$   in the $\pi^\pm h_c$ invariant mass spectrum in the electron-positron collisions $e^+e^- \to \pi^+\pi^- h_c$ \cite{BES1309}. Now the $Z^\pm_c(4020)$ and $Z^\pm_c(4025)$ are listed in {\it The Review of Particle Physics}  as the same particle  \cite{PDG}.
In 2014, the LHCb collaboration performed a four-dimensional fit of the scattering  amplitude for the decay $B^0\to\psi'\pi^-K^+$  in proton-proton collisions, and  obtained  the first independent confirmation of
the charmonium-like resonance  $Z^-_c(4430)$
and determined  its quantum numbers to be  $J^P=1^+$  \cite{LHCb-1404}. In 2017, the BESIII collaboration established the charmonium-like resonance
 $Z_c(3900)$'s quantum numbers
to be $J^P = 1^+$ \cite{BES-Zc3900-JP}.

There have been several possible explanations for the exotic states $Z_c(3900)$ and $Z_c(4020)$, such as
the molecular states (from the heavy quark symmetries \cite{ZhaoQ-3900,GuoFK-3900-4020}, the QCD sum rules \cite{HuangMQ-3900,WangZG-4-quark-mole},
the light-front quark model \cite{KeHW-3900},  the one-pion exchange model \cite{HeJun-4020}, and the phenomenological Lagrangian approach \cite{DongYB-3900-4020}),
 the tetraquark states (from the diquark model with the effective Hamiltonian \cite{Maiani-3900}, the QCD sum rules \cite{WangHuangTao,QiaoCF-3900-4020,Wang-4020-CTP,WangEPJC-1601}, and
 the potential model  \cite{PingJL-4020}),  the triangle singularities  (in the rescattering amplitudes) \cite{LiuXH-3900}, the threshold effects \cite{ChenDY-3900}, etc.

We can tentatively assign the hidden-charm resonances $Z_c(3900)$ and $Z_c(4430)$ as the ground state  tetraquark state and the first radially excited
 tetraquark state respectively  considering
 the similar decays, $Z_c(3900)^\pm \to J/\psi\pi^\pm$, $Z_c(4430)^\pm \to \psi^\prime\pi^\pm$,
and the almost equal  energy  gaps $M_{Z_c(4430)}-M_{Z_c(3900)}=591\,\rm{MeV}$, $M_{\psi^\prime}-M_{J/\psi}=589\,\rm{MeV}$ \cite{Z4430-1405,Nielsen-1401}.
In Ref.\cite{Wang4430}, we adopt the  method  invented  in Ref.\cite{Baxi-G} for the conventional  quarkonium  to study the  $Z^\pm_c(3900)$ as the ground state axialvector tetraquark state and the $Z^\pm_c(4430)$ as the first radially excited  axialvector  tetraquark state respectively,  and employ  the energy scale formula
$\mu=\sqrt{M^2_{X/Y/Z}-(2{\mathbb{M}}_c)^2}$ to select the best energy scales of the  spectral densities at the QCD side of the QCD sum rules with the effective
(or constituent)  charm quark mass   ${\mathbb{M}}_c$  \cite{Wang-tetra-formula}. In Ref.\cite{Azizi-4430}, this subject is studied with the QCD sum rules adopting  another parameter system. In Refs.\cite{Wang-4020-CTP,WangEPJC-1601}, we observe that we can assign the $Z_c(4020/4025)$  to be the ground state axialvector $[uc]_A[\bar{d}\bar{c}]_A$ tetraquark state with the quantum numbers $J^{PC}=1^{+-}$ according to the QCD sum rules calculations. If the $Z_c(4600)$ is the first radial excitation of the hidden-charm tetraquark candidate $Z_c(4020/4025)$, its preferred decay mode is $Z_c(4600)\to \psi^\prime \pi$ rather than $Z_c(4600)\to J/\psi \pi$.

 In this paper, we intend to perform a detailed and updated analysis of the ground states  and the first radially excited states of the charged  hidden-charm tetraquark states with the QCD sum rules, and explore the possible assignments  of the $Z_c(4600)$ state in the  scenario of the axialvector hidden-charm tetraquark states with the quantum numbers $J^{PC}=1^{+-}$.

The paper is arranged as follows: in Sect.2 we  obtain the analytical expressions of the  QCD sum rules for  the hidden-charm axialvector tetraquark states $Z_c$;
in Sect.3  we  provide the numerical results for the masses and pole residues of the $Z_c$ states and detailed discussions; in Sect.4 we reach  the conclusion.

\section{The QCD sum rules for the axialvector tetraquark states}
In the following, let us  write down  the two-point Green functions  (or correlation functions)   $\Pi_{\mu\nu}(p)$ and $\Pi_{\mu\nu\alpha\beta}(p)$ as the first step,
\begin{eqnarray}
\Pi_{\mu\nu}(p)&=&i\int d^4x e^{ip \cdot x} \langle0|T\Big\{J_\mu(x)J_{\nu}^{\dagger}(0)\Big\}|0\rangle \, ,\nonumber\\
\Pi_{\mu\nu\alpha\beta}(p)&=&i\int d^4x e^{ip \cdot x} \langle0|T\Big\{J_{\mu\nu}(x)J_{\alpha\beta}^{\dagger}(0)\Big\}|0\rangle \, ,
\end{eqnarray}
where the four-quark current operators  $J_\mu(x)=J^1_\mu(x)$, $J^2_\mu(x)$, $J^3_\mu(x)$,
\begin{eqnarray}
J^1_\mu(x)&=&\frac{\varepsilon^{ijk}\varepsilon^{imn}}{\sqrt{2}}\Big[u^{Tj}(x)C\gamma_5c^k(x) \bar{d}^m(x)\gamma_\mu C \bar{c}^{Tn}(x)-u^{Tj}(x)C\gamma_\mu c^k(x)\bar{d}^m(x)\gamma_5C \bar{c}^{Tn}(x) \Big] \, ,\nonumber\\
J_\mu^2(x)&=&\frac{\varepsilon^{ijk}\varepsilon^{imn}}{\sqrt{2}}\Big[u^{Tj}(x)C\sigma_{\mu\nu}\gamma_5 c^k(x)\bar{d}^m(x)\gamma^\nu C \bar{c}^{Tn}(x)-u^{Tj}(x)C\gamma^\nu c^k(x)\bar{d}^m(x)\gamma_5\sigma_{\mu\nu} C \bar{c}^{Tn}(x) \Big] \, , \nonumber\\
J_\mu^3(x)&=&\frac{\varepsilon^{ijk}\varepsilon^{imn}}{\sqrt{2}}\Big[u^{Tj}(x)C\sigma_{\mu\nu} c^k(x)\bar{d}^m(x)\gamma_5\gamma^\nu C \bar{c}^{Tn}(x)+u^{Tj}(x)C\gamma^\nu \gamma_5c^k(x)\bar{d}^m(x) \sigma_{\mu\nu} C \bar{c}^{Tn}(x) \Big] \, , \nonumber \\
J_{\mu\nu}(x)&=&\frac{\varepsilon^{ijk}\varepsilon^{imn}}{\sqrt{2}}\Big[u^{Tj}(x) C\gamma_\mu c^k(x) \bar{d}^m(x) \gamma_\nu C \bar{c}^{Tn}(x)  -u^{Tj}(x) C\gamma_\nu c^k(x) \bar{d}^m(x) \gamma_\mu C \bar{c}^{Tn}(x) \Big] \, ,
\end{eqnarray}
  the superscripts $i$, $j$, $k$, $m$ and $n$ are color indexes with values obeying the antisymmetric tensor $\varepsilon$, the charge conjugation matrix $C=i\gamma^2\gamma^0$.
If we preform  charge conjugation (and parity) transform $\widehat{C}$ (and $\widehat{P}$), the axialvector current operators $J_\mu(x)$ and tensor current operator $J_{\mu\nu}(x)$ have the following properties,
\begin{eqnarray}
\widehat{C}J_{\mu}(x)\widehat{C}^{-1}&=&- J_\mu(x) \, , \nonumber\\
\widehat{C}J_{\mu\nu}(x)\widehat{C}^{-1}&=&- J_{\mu\nu}(x) \, , \nonumber\\
\widehat{P}J_{\mu}(x)\widehat{P}^{-1}&=&- J^\mu(\tilde{x}) \, , \nonumber\\
\widehat{P}J_{\mu\nu}(x)\widehat{P}^{-1}&=& J^{\mu\nu}(\tilde{x}) \, ,
\end{eqnarray}
where the coordinates $x^\mu=(t,\vec{x})$ and $\tilde{x}^\mu=(t,-\vec{x})$.

  The diquark  operators $\varepsilon^{ijk}q^{T}_j C\Gamma Q_k$ in the attractive color-antitriplet $\bar{3}_c$ channel have  five spinor structures, where $C\Gamma=C\gamma_5$, $C$, $C\gamma_\mu \gamma_5$,  $C\gamma_\mu $ and $C\sigma_{\mu\nu}$ or $C\sigma_{\mu\nu}\gamma_5$ correspond to the  scalar, pseudoscalar,
  vector, axialvector  and  tensor  diquark operators, respectively.
  The favorable  quark-quark correlations  are the scalar diquark  and axialvector  diquark  in the color-antitriplet $\bar{3}_c$ channel  from the  QCD sum rules \cite{WangDiquark}.
 If we introduce  a relative  P-wave between the light quark and heavy quark, we can obtain the pseudoscalar diquark operator $\varepsilon^{ijk}q^{T}_j C\gamma_5 \underline{\gamma_5}Q_k$ and vector diquark operator  $\varepsilon^{ijk}q^{T}_jC\gamma_\mu\underline{\gamma_5} Q_k$  without introducing the additional P-wave explicitly, as multiplying a $\gamma_5$ can change the parity, the P-wave effect is embodied in the underlined $\gamma_5$.
   The pseudoscalar diquark states and vector diquark states (or the P-wave diquark states)   have larger masses compared   with the scalar diquark states and axialvector  diquark states, we would like to choose the  scalar diquark  and axialvector  diquark to construct the four-quark  current operators to interpolate the lower tetraquark states.

The  tensor heavy  diquark operators $\varepsilon^{abc}q^{T}_b(x)C\sigma_{\mu\nu}\gamma_5Q_c(x)$ and $\varepsilon^{abc}q^{T}_b(x)C\sigma_{\mu\nu}Q_c(x)$ have both axialvector constituents  and vector constituents,
\begin{eqnarray}
\widehat{P}\varepsilon^{abc}q^{T}_b(x)C\sigma_{jk}\gamma_5Q_c(x)\widehat{P}^{-1}&=&+\varepsilon^{abc}q^{T}_b(\tilde{x})C\sigma_{jk}\gamma_5Q_c(\tilde{x}) \, , \nonumber\\
\widehat{P}\varepsilon^{abc}q^{T}_b(x)C\sigma_{0j} Q_c(x)\widehat{P}^{-1}&=&+\varepsilon^{abc}q^{T}_b(\tilde{x})C\sigma_{0j} Q_c(\tilde{x}) \, , \nonumber\\
\widehat{P}\varepsilon^{abc}q^{T}_b(x)C\sigma_{0j}\gamma_5Q_c(x)\widehat{P}^{-1}&=&-\varepsilon^{abc}q^{T}_b(\tilde{x})C\sigma_{0j}\gamma_5Q_c(\tilde{x}) \, , \nonumber\\
\widehat{P}\varepsilon^{abc}q^{T}_b(x)C\sigma_{jk} Q_c(x)\widehat{P}^{-1}&=&-\varepsilon^{abc}q^{T}_b(\tilde{x})C\sigma_{jk} Q_c(\tilde{x})\, ,
\end{eqnarray}
where the space indexes $j$, $k=1$, $2$, $3$.
The tensor diquark operators also play an important role in constructing the tetraquark current operators \cite{Wang-Y4140-Y4274}.
 We  multiply the tensor diquark (antidiquark) operators with the axialvector or vector antidiquark (diquark) operators so as to project out the axialvector constituents and  vector constituents  to construct the four-quark current operators $J^2_\mu(x)$ and $J^3_\mu(x)$.  Thereafter,  we will use the $\tilde{V}$ and $\tilde{A}$ to represent  the vector  constituent and axialvector  constituent  of the tensor diquark operators, respectively.

  The four-quark current operators $J^1_\mu(x)$, $J^2_\mu(x)$ and $J^3_\mu(x)$ couple potentially to the
$[uc]_S[\bar{d}\bar{c}]_A-[uc]_A[\bar{d}\bar{c}]_S$ type,
$[uc]_{\tilde{A}}[\bar{d}\bar{c}]_A-[uc]_A[\bar{d}\bar{c}]_{\tilde{A}}$ type and $[uc]_{\tilde{V}}[\bar{d}\bar{c}]_V+[uc]_V[\bar{d}\bar{c}]_{\tilde{V}}$ type
 axialvector hidden-charm  tetraquark states with the spin-parity-charge-conjugation $J^{PC}=1^{+-}$, respectively. While the current operator  $J_{\mu\nu}(x)$ couples potentially to  both
 the $[uc]_A[\bar{d}\bar{c}]_A$ type axialvector tetraquark state with $J^{PC}=1^{+-}$ and vector tetraquark state with $J^{PC}=1^{--}$.  Thereafter, we will not distinguish the negative   or positive  electric-charge of the $Z_c$ tetraquark states, as they have degenerate masses.

We insert  a complete set of hadron states which have nonvanishing couplings  with the four-quark current operators $J_\mu(x)$ and $J_{\mu\nu}(x)$ into the Green functions $\Pi_{\mu\nu}(p)$ and $\Pi_{\mu\nu\alpha\beta}(p)$ to get  the hadronic  representation
\cite{SVZ79,Reinders85}. Then we separate  the ground state
  axialvector and vector tetraquark state contributions from other contributions, such as the higher excited tetraquark states and continuum states, to obtain  the results,
\begin{eqnarray}
\Pi_{\mu\nu}(p)&=&\frac{\lambda_{Z}^2}{m^2_{Z}-p^2}\left(-g_{\mu\nu } +\frac{p_\mu p_{\nu}}{p^2}\right) +\cdots  \nonumber\\
&=&\Pi_Z(p^2)\left(-g_{\mu\nu } +\frac{p_\mu p_{\nu}}{p^2}\right) +\cdots \, ,\\
\Pi_{\mu\nu\alpha\beta}(p)&=&\frac{\widetilde{\lambda}_{ Z}^2}{m_{Z}^2-p^2}\left(p^2g_{\mu\alpha}g_{\nu\beta} -p^2g_{\mu\beta}g_{\nu\alpha} -g_{\mu\alpha}p_{\nu}p_{\beta}-g_{\nu\beta}p_{\mu}p_{\alpha}+g_{\mu\beta}p_{\nu}p_{\alpha}+g_{\nu\alpha}p_{\mu}p_{\beta}\right) \nonumber\\
&&+\frac{\widetilde{\lambda}_{ Y}^2}{m_{Y}^2-p^2}\left( -g_{\mu\alpha}p_{\nu}p_{\beta}-g_{\nu\beta}p_{\mu}p_{\alpha}+g_{\mu\beta}p_{\nu}p_{\alpha}+g_{\nu\alpha}p_{\mu}p_{\beta}\right) +\cdots \nonumber\\
&=&\Pi_Z(p^2)\left(p^2g_{\mu\alpha}g_{\nu\beta} -p^2g_{\mu\beta}g_{\nu\alpha} -g_{\mu\alpha}p_{\nu}p_{\beta}-g_{\nu\beta}p_{\mu}p_{\alpha}+g_{\mu\beta}p_{\nu}p_{\alpha}+g_{\nu\alpha}p_{\mu}p_{\beta}\right) \nonumber\\
&&+\Pi_Y(p^2)\left( -g_{\mu\alpha}p_{\nu}p_{\beta}-g_{\nu\beta}p_{\mu}p_{\alpha}+g_{\mu\beta}p_{\nu}p_{\alpha}+g_{\nu\alpha}p_{\mu}p_{\beta}\right) \, ,
\end{eqnarray}
where the $Z$ represents the axialvector tetraquark states, the $Y$ represents the vector tetraquark states, the    $\lambda_{Z}$, $\widetilde{\lambda}_{Z}$ and $\widetilde{\lambda}_{Y}$ are the pole residues or current-tetraquark coupling constants,
\begin{eqnarray}
\langle 0|J_\mu(0)|Z_c(p)\rangle&=&\lambda_{Z}\, \varepsilon_\mu\, ,\nonumber\\
  \langle 0|J_{\mu\nu}(0)|Z_c(p)\rangle &=& \widetilde{\lambda}_{Z} \, \varepsilon_{\mu\nu\alpha\beta} \, \varepsilon^{\alpha}p^{\beta}\, , \nonumber\\
 \langle 0|J_{\mu\nu}(0)|Y(p)\rangle &=& \widetilde{\lambda}_{Y} \left(\varepsilon_{\mu}p_{\nu}-\varepsilon_{\nu}p_{\mu} \right)\, ,
\end{eqnarray}
the antisymmetric tensor $\varepsilon_{0123}=-1$, the  $\varepsilon_\mu(\lambda,p)$ are the polarization vectors of the axialvector   and vector tetraquark states satisfy the summation formula,
 \begin{eqnarray}
\sum_{\lambda}\varepsilon^*_{\mu}(\lambda,p)\varepsilon_{\nu}(\lambda,p)&=&-g_{\mu\nu}+\frac{p_\mu p_\nu}{p^2} \, .
 \end{eqnarray}

The diquark-antidiquark type four-quark  current operators $J_\mu(x)$ and $J_{\mu\nu}(x)$ couple potentially to the diquark-antidiquark type hidden-charm tetraquark states.
We can perform Fierz rearrangements  to those  currents    in both the spinor space and color space to obtain a series of  color-singlet-color-singlet (or meson-meson) type current operators, for example,
\begin{eqnarray}
J_{\mu}^1(x)&=&\frac{1}{2\sqrt{2}}\Big\{\,i\bar{c}(x)i\gamma_5 c(x)\,\bar{d}(x)\gamma^\mu u(x)-i\bar{c}(x) \gamma^\mu c(x)\,\bar{d}(x)i\gamma_5 u(x)+\bar{c}(x) u(x)\,\bar{d}(x)\gamma^\mu\gamma_5 c(x)  \nonumber\\
&&  -\bar{c}(x) \gamma^\mu \gamma_5u(x)\,\bar{d}(x)c(x)- i\bar{c}(x)\gamma_\nu\gamma_5c(x)\, \bar{d}(x)\sigma^{\mu\nu}u(x)+i\bar{c}(x)\sigma^{\mu\nu}c(x)\, \bar{d}(x)\gamma_\nu\gamma_5u(x)\nonumber\\
&&- i \bar{c}(x)\sigma^{\mu\nu}\gamma_5u(x)\,\bar{d}(x)\gamma_\nu c(x)+i\bar{c}(x)\gamma_\nu u(x)\, \bar{d}(x)\sigma^{\mu\nu}\gamma_5c(x)   \,\Big\} \, ,
\end{eqnarray}
while the  constituents, such as $\bar{c}(x)i\gamma_5 c(x)\,\bar{d}(x)\gamma^\mu u(x)$, $\bar{c}(x) \gamma^\mu c(x)\,\bar{d}(x)i\gamma_5 u(x)$, etc,  couple potentially  to the  meson-meson type scattering  states  or tetraquark molecular states.

However, we should be careful  in performing  the Fierz rearrangements,  the rearrangements in the spinor space and color  space  are quite non-trivial, the scenarios of the diquark-antidiquark type tetraquark states and meson-meson type molecular states are quite  different.

According to the arguments of Selem and Wilczek,  a diquark-antidiquark type tetraquark can  be plausibly described by two diquarks  trapped  in a double potential well, the two potential wells are  separated apart by a barrier \cite{Wilczek-diquark}. At long
distances, the diquark and antiquark serve as  point color charges respectively, and attract each other strongly  just like in the  quark and antiquark bound states. However, when the two diquarks approach each other,  the attractions between quark and antiquark in different diquarks decrease
the bound energy  of the diquarks and tend to destroy the diquarks. Those effects (beyond the naive one-gluon exchange force) increase
when the distance between the diquark and antidiquark  decreases,  and a repulsive interaction  between
the diquark and antidiquark emerges, if  large enough,  it
will lead to  a barrier between the diquark and antidiquark \cite{Polosa-diquark}. The two  potential wells
which are separated apart  by a barrier can give successful  descriptions of the diquark-antidiquark type   tetraquark states \cite{Polosa-diquark}.

While in the dynamical picture of the tetraquark states, the large spatial separation between the diquark and antidiquark leads  to small wave-function overlap between the quark-antiquark pair \cite{Brodsky-PRL}, the rearrangements in the spinor space and color  space are highly suppressed.

In practical calculations, it is  difficult to account for the non-local effects between the diquark and antidiquark pair in the four-quark  currents $J_\mu(x)$ and $J_{\mu\nu}(x)$ directly,
for example, the current $J^1_\mu(x)$ can be modified to
\begin{eqnarray}
J^1_\mu(x,\epsilon)&=&\frac{\varepsilon^{ijk}\varepsilon^{imn}}{\sqrt{2}}\Big[u^{Tj}(x)C\gamma_5c^k(x) \bar{d}^m(x+\epsilon)\gamma_\mu C \bar{c}^{Tn}(x+\epsilon)-u^{Tj}(x)C\gamma_\mu c^k(x)\bar{d}^m(x+\epsilon)\nonumber\\
&&\gamma_5C \bar{c}^{Tn}(x+\epsilon) \Big] \, ,
\end{eqnarray}
to account for the non-locality by adding a finite $\epsilon$, however,  it is very difficult to deal with the finite $\epsilon$ both at the hadron side and at the QCD side in a consistent way, we  expend the current $J^1_\mu(x,\epsilon)$  in terms of Taylor series of $\epsilon$,
\begin{eqnarray}
J^1_\mu(x,\epsilon)&=&J^1_\mu(x,0)+ \frac{\partial J^1_\mu(x,\epsilon)}{\partial\epsilon^\alpha}\mid_{\epsilon=0} \epsilon^\alpha+\frac{1}{2} \frac{\partial^2 J^1_\mu(x,\epsilon)}{\partial\epsilon^\alpha\partial\epsilon^\beta}\mid_{\epsilon=0} \epsilon^\alpha\epsilon^\beta+\cdots\, ,
\end{eqnarray}
then  expend the correlation function $\Pi_{\mu\nu}(p)$ also in terms of Taylor series of $\epsilon$,
\begin{eqnarray}
\Pi_{\mu\nu}(p)&=&\Pi_{\mu\nu}\left(\mathcal{O}(\epsilon^0)\right)+\Pi_{\mu\nu}\left(\mathcal{O}(\epsilon^1)\right)+\Pi_{\mu\nu}\left(\mathcal{O}(\epsilon^2)\right)+\cdots\, , \end{eqnarray}
where the components $\Pi_{\mu\nu}\left(\mathcal{O}(\epsilon^i)\right)$ with $i=0$, $1$, $2$, $\cdots$ stand for  the contributions of the order $\mathcal{O}(\epsilon^i)$. In this article, we study the  leading order contributions $J^1_\mu(x)=J^1_\mu(x,0)$ and $\Pi_{\mu\nu}(p)=\Pi_{\mu\nu}\left(\mathcal{O}(\epsilon^0)\right)$, the effects beyond the leading order frustrate   the Fierz rearrangements of the  diquark-antidiquark type currents  into a series of  color-singlet-color-singlet (meson-meson) type currents freely.

We can use the Feynman diagram drawn in Fig.1 to describe the lowest order  contributions in the correlation functions for the  diquark-antidiquark type four-quark currents, and use the Feynman diagrams drawn in Fig.2 to describe   the corresponding lowest order  contributions in the correlation functions for the
color-singlet-color-singlet type four-quark  currents. The Feynman diagram drawn in Fig.1 cannot be factorized into the two Feynman diagrams drawn in Fig.2 freely due to the
barrier (or spatial separation) between the diquark and antidiquark \cite{Polosa-diquark,Brodsky-PRL}. When a quark (antiquark) in the diquark (antidiquark) penetrates the barrier, the Feynman diagram drawn in Fig.1 is  factorizable in color space. In this case, the non-factorizable diagrams start at the order $\mathcal{O}(\alpha_s^2)$ \cite{Melikhov-factor}.

In Ref.\cite{Melikhov-factor}, Lucha,  Melikhov, and Sazdjiand argue that the diquark-antidiquark type four-quark currents can
be changed into the color-singlet-color-singlet (meson-meson) type currents through Fierz transformation, the Feynman diagrams which make contributions   to the quark-gluon operators of the
order $\mathcal{O}(\alpha_s^0)$ and $\mathcal{O}(\alpha_s)$ in accomplishing the operator product expansion are  factorizable and are canceled out by the contributions of the two-meson  scattering states at the phenomenological side, furthermore, the factorizable parts (in color space) of the Feynman diagrams of the order $\mathcal{O}(\alpha_s^2)$ are also canceled out by the contributions come from the two-meson scattering  states (or more precisely, the free two-meson states),
 the relevant non-factorizable contributions  start at the order  $\mathcal{O}(\alpha_s^2)$. We do not agree with their viewpoint, as there exists  a repulsive barrier \cite{Wilczek-diquark,Polosa-diquark} or
 a large spatial separation \cite{Brodsky-PRL}, which are embodied in the non-local effects, to prevent  performing  the Fierz transformation freely, although at the present time we cannot take into account the   non-local effects in the QCD sum rules, and have to take the  leading order approximations $J^1_\mu(x)=J^1_\mu(x,0)$ and $\Pi_{\mu\nu}(p)=\Pi_{\mu\nu}\left(\mathcal{O}(\epsilon^0)\right)$. Our viewpoint is that the relevant contributions begin at the order $\mathcal{O}(\alpha_s^0)$, it is not necessary  to perform or it is very  difficult to perform  the Fierz transformation to separate the factorizable and non-factorizable
 contributions in the color space, we should take into account both the factorizable and nonfactorizable  Feynman diagrams for the diquark-antidiquark type currents.

\begin{figure}
 \centering
  \includegraphics[totalheight=3cm,width=5cm]{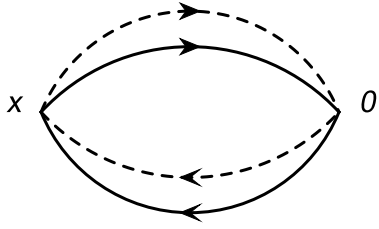}
 \caption{ The  Feynman diagram  of the lowest order  contributions for the diquark-antidiquark type currents, where the solid lines represent the light quarks and dashed lines represent the heavy quarks.  }
\end{figure}

\begin{figure}
 \centering
  \includegraphics[totalheight=3cm,width=10cm]{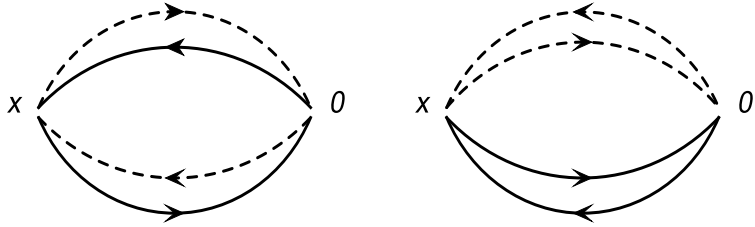}
 \caption{ The  Feynman diagrams  of the lowest order  contributions for the color-singlet-color-singlet type currents, where the solid lines represent the light quarks and dashed lines represent the heavy quarks.  }
\end{figure}

When the quark or antiquark penetrates the barrier, we can perform the Fierz rearrangements, and study the effects  of the scattering states.
 Now let us begin to explore  the  contributions of the meson-meson type scattering states (in other words, the two-meson loops) to the Green function $\Pi_{\mu\nu}(p)$ for the four-quark current $J^1_\mu(x)$ as a representative example,
\begin{eqnarray}
\Pi_{\mu\nu}(p)&=&-\frac{\widehat{\lambda}_{Z}^{2}}{ p^2-\widehat{M}_{Z}^2}\widetilde{g}_{\mu\nu}(p)-\frac{\widehat{\lambda}_{Z}}{p^2-\widehat{M}_{Z}^2}\widetilde{g}_{\mu\alpha}(p)
 \Sigma_{DD^*}(p) \widetilde{g}^{\alpha\beta}(p) \widetilde{g}_{\beta\nu}(p)\frac{\widehat{\lambda}_{Z}}{p^2-\widehat{M}_{Z}^2} \nonumber \\
 &&-\frac{\widehat{\lambda}_{Z}}{p^2-\widehat{M}_{Z}^2}\widetilde{g}_{\mu\alpha}(p)
 \Sigma_{J/\psi\pi}(p) \widetilde{g}^{\alpha\beta}(p) \widetilde{g}_{\beta\nu}(p)\frac{\widehat{\lambda}_{X/Z}}{p^2-\widehat{M}_{Z}^2}  +\cdots \, ,\nonumber \\
 &=&-\frac{\widehat{\lambda}_{Z}^{2}}{ p^2-\widehat{M}_{Z}^2-\Sigma_{DD^*}(p)-\Sigma_{J/\psi\pi}(p)+\cdots}\widetilde{g}_{\mu\nu}(p)+\cdots \, ,
\end{eqnarray}
where
\begin{eqnarray}
\Sigma_{DD^*}(p)&=&i\int~{d^4q\over(2\pi)^4}\frac{G^2_{ZDD^*}}{\left[q^2-M_{D}^2\right]\left[ (p-q)^2-M_{D^*}^2\right]} \, , \nonumber\\
\Sigma_{J/\psi \pi}(p)&=&i\int~{d^4q\over(2\pi)^4}\frac{G^2_{ZJ/\psi \pi}}{\left[q^2-M_{J/\psi}^2\right]\left[ (p-q)^2-M_{\pi}^2\right]} \, ,
\end{eqnarray}
$\widetilde{g}_{\mu\nu}(p)=-g_{\mu\nu}+\frac{p_{\mu}p_{\nu}}{p^2}$, the $G_{Z DD^*}$ and $G_{Z J/\psi\pi}$ are the hadronic coupling constants. We resort to the bare  quantities  $\widehat{\lambda}_{Z}$ and $\widehat{M}_{Z}$ so as  to absorb the divergent terms which appear in the integrals in calculating  the self-energies $\Sigma_{DD^*}(p)$, $\Sigma_{J/\psi \pi}(p)$, etc.
The self-energies after renormalization  result in  a finite energy-dependent width to modify the dispersion relation,
\begin{eqnarray}
\Pi_{\mu\nu}(p) &=&-\frac{\lambda_{Z}^{2}}{ p^2-M_{Z}^2+i\sqrt{p^2}\Gamma(p^2)}\widetilde{g}_{\mu\nu}(p)+\cdots \, ,
 \end{eqnarray}
the experimental value of the total decay  width $\Gamma_{Z_c(3900)}(M_Z^2)=(46\pm 10\pm 20) \,\rm{MeV}$ \cite{BES3900} (or $(28.2\pm 2.6)\, \rm{MeV}$ \cite{PDG}),
 the zero width approximation in  the  spectral densities at  the phenomenological side are approved  reasonable  \cite{WangZG-Zc4200}.  In this paper, we neglect the contributions of the meson-meson type scattering states or the two-meson loops, the predictions are still robust.

We calculate all the Feynman diagrams in performing the operator product expansion to obtain the QCD spectral representation of the  Green functions (or correlation functions) $\Pi_{\mu\nu}(p)$ and $\Pi_{\mu\nu\alpha\beta}(p)$.
 In analytical  calculations,  we take into account the vacuum condensates (by selecting   the quark-gluon operators  of the orders $\mathcal{O}( \alpha_s^{k})$ with $k\leq 1$) up to dimension $10$ consistently and factorize the higher dimensional condensates into the lower dimensional condensates by assuming vacuum saturation.
 After getting the analytical expressions of the Green  functions at the quark-gluon level, we obtain the  spectral representation via dispersion relation.
 Now  we  match the hadronic representation with the QCD representation of the Green functions (or correlation functions)  $\Pi_Z(p^2)$
 below the continuum threshold parameters $s_0$ and carry out the  Borel transformation  in regard to
 $P^2=-p^2$ to get  the  QCD sum rules:
\begin{eqnarray}\label{QCDSR-I1}
\lambda^2_{Z}\, \exp\left(-\frac{M^2_{Z}}{T^2}\right)= \int_{4m_c^2}^{s_0} ds\, \rho(s) \, \exp\left(-\frac{s}{T^2}\right) \, ,
\end{eqnarray}
\begin{eqnarray}
\rho(s)&=&\rho_{0}(s)+\rho_{3}(s) +\rho_{4}(s)+\rho_{5}(s)+\rho_{6}(s)+\rho_{7}(s) +\rho_{8}(s)+\rho_{10}(s)\, ,
\end{eqnarray}
 $\lambda_Z=\widetilde{\lambda}_Z M_Z$, the $T^2$ is the Borel parameter,  the subscripts $i$ in the components of the QCD spectral
 densities $\rho_{i}(s)$ represent the dimensions of the vacuum condensates,
\begin{eqnarray}
\rho_{3}(s)&\propto& \langle\bar{q}q\rangle\, ,\nonumber\\
\rho_{4}(s)&\propto& \langle\frac{\alpha_{s}GG}{\pi}\rangle\, ,\nonumber\\
\rho_{5}(s)&\propto& \langle\bar{q}g_s\sigma Gq\rangle\, ,\nonumber\\
\rho_{6}(s)&\propto& \langle\bar{q}q\rangle^2\, ,\,4\pi\alpha_s\langle\bar{q}q\rangle^2\, ,\nonumber\\
\rho_{7}(s)&\propto& \langle\bar{q}q\rangle\langle\frac{\alpha_{s}GG}{\pi}\rangle\, ,\nonumber\\
\rho_{8}(s)&\propto& \langle\bar{q}q\rangle\langle\bar{q}g_s\sigma Gq\rangle\, ,\nonumber\\
\rho_{10}(s)&\propto&  \langle\bar{q}g_s\sigma Gq\rangle^2\, ,\, \langle\bar{q}q\rangle^2\langle\frac{\alpha_{s}GG}{\pi}\rangle\, .
\end{eqnarray}
We neglect the cumbersome analytical expressions of the  spectral densities at the quark-gluon level   in saving printed pages.
  We can refer to Ref.\cite{WangHuangTao} for the technical details in calculating the Feynman diagrams. On the other hand,
 we can refer to Refs.\cite{WangHuangTao,WangEPJC-1601} for the explicit expressions of the spectral densities at the quark-gluon level for the axialvector current $J^1_\mu(x)$ and tensor current $J_{\mu\nu}(x)$.
In this work, we recalculate those QCD spectral densities, and use the formula $t^a_{ij}t^a_{mn}=-\frac{1}{6}\delta_{ij}\delta_{mn}+\frac{1}{2}\delta_{jm}\delta_{in}$ with $t^a=\frac{\lambda^a}{2}$ to deal with the higher  dimensional vacuum condensates, where the $\lambda^a$ is the Gell-Mann matrix. This routine leads to slight  but neglectful   differences compared to
  the old calculations.  For the currents $J^2_\mu(x)$ and $J^3_\mu(x)$, we neglect the tiny contributions of the  $4\pi\alpha_s\langle\bar{q}q\rangle^2$, which originate from the operators  like $\langle\bar{q}_j\gamma_{\mu}q_ig_s D_\nu G^a_{\alpha\beta}t^a_{mn}\rangle$.

  We derive  Eq.\eqref{QCDSR-I1} in regard to  $\tau=\frac{1}{T^2}$, then  reach   the QCD sum rules for the tetraquark  masses by eliminating the
 pole residues  $\lambda_{Z}$ through a fraction,
 \begin{eqnarray}\label{QCDSR-I2}
M_{Z}^2&=&- \frac{\int_{4m_c^2}^{s_0} ds \, \frac{d}{d \tau }\,\rho(s)e^{-\tau s}}{\int_{4m_c^2}^{s_0} ds \rho(s)e^{-\tau s}}\, .
\end{eqnarray}
Thereafter, we will refer the QCD sum rules in Eq.\eqref{QCDSR-I1} and Eq.\eqref{QCDSR-I2} as QCDSR I.

If we take into account the contributions of the first radially excited tetraquark states $Z_c^\prime$ in the hadronic representation, we can obtain the QCD sum rules,
\begin{eqnarray}\label{QCDSR-II}
\lambda^2_{Z}\, \exp\left(-\frac{M^2_{Z}}{T^2}\right)+\lambda^2_{Z^\prime}\, \exp\left(-\frac{M^2_{Z^\prime}}{T^2}\right)&=& \int_{4m_c^2}^{s_0^\prime} ds\, \rho(s) \, \exp\left(-\frac{s}{T^2}\right) \, ,
\end{eqnarray}
where the $s_0^\prime$ is continuum threshold parameter, then we   introduce the notations $\tau=\frac{1}{T^2}$, $D^n=\left( -\frac{d}{d\tau}\right)^n$, and  resort to the subscripts $1$ and $2$ to represent  the ground  state tetraquark state $Z_c$ and the first radially excited tetraquark  state $Z_c^\prime$ respectively for simplicity.
 We rewrite the   QCD sum rules as
\begin{eqnarray}\label{QCDSR-II-re}
\lambda_1^2\exp\left(-\tau M_1^2 \right)+\lambda_2^2\exp\left(-\tau M_2^2 \right)&=&\Pi_{QCD}(\tau) \, ,
\end{eqnarray}
here we introduce the subscript $QCD$ to represent the QCD representation.
We derive  the QCD sum rules in Eq.\eqref{QCDSR-II-re} in regard  to $\tau$ to get
\begin{eqnarray}\label{QCDSR-II-Dr}
\lambda_1^2M_1^2\exp\left(-\tau M_1^2 \right)+\lambda_2^2M_2^2\exp\left(-\tau M_2^2 \right)&=&D\Pi_{QCD}(\tau) \, .
\end{eqnarray}
From Eqs.\eqref{QCDSR-II-re}-\eqref{QCDSR-II-Dr}, we obtain the QCD sum rules,
\begin{eqnarray}\label{QCDSR-II-Residue}
\lambda_i^2\exp\left(-\tau M_i^2 \right)&=&\frac{\left(D-M_j^2\right)\Pi_{QCD}(\tau)}{M_i^2-M_j^2} \, ,
\end{eqnarray}
where the indexes $i \neq j$.
Let us derive   the QCD sum rules in Eq.\eqref{QCDSR-II-Residue} in regard  to $\tau$ to get
\begin{eqnarray}
M_i^2&=&\frac{\left(D^2-M_j^2D\right)\Pi_{QCD}(\tau)}{\left(D-M_j^2\right)\Pi_{QCD}(\tau)} \, , \nonumber\\
M_i^4&=&\frac{\left(D^3-M_j^2D^2\right)\Pi_{QCD}(\tau)}{\left(D-M_j^2\right)\Pi_{QCD}(\tau)}\, .
\end{eqnarray}
 The squared masses $M_i^2$ satisfy the  equation,
\begin{eqnarray}
M_i^4-b M_i^2+c&=&0\, ,
\end{eqnarray}
where
\begin{eqnarray}
b&=&\frac{D^3\otimes D^0-D^2\otimes D}{D^2\otimes D^0-D\otimes D}\, , \nonumber\\
c&=&\frac{D^3\otimes D-D^2\otimes D^2}{D^2\otimes D^0-D\otimes D}\, , \nonumber\\
D^j \otimes D^k&=&D^j\Pi_{QCD}(\tau) \,  D^k\Pi_{QCD}(\tau)\, ,
\end{eqnarray}
the indexes $i=1,2$ and $j,k=0,1,2,3$.
Finally we solve above equation analytically to obtain two solutions \cite{Baxi-G},
\begin{eqnarray}\label{QCDSR-II-M1}
M_1^2&=&\frac{b-\sqrt{b^2-4c} }{2} \, ,
\end{eqnarray}
\begin{eqnarray}\label{QCDSR-II-M2}
M_2^2&=&\frac{b+\sqrt{b^2-4c} }{2} \, .
\end{eqnarray}
From now on, we will denote  the QCD sum rules in Eq.\eqref{QCDSR-II} and Eqs.\eqref{QCDSR-II-M1}-\eqref{QCDSR-II-M2} as QCDSR II.
In calculations, we observe that if we specify the energy scales of the  spectral densities in the QCD representation,   only one solution satisfies the energy scale formula $\mu=\sqrt{M^2_{X/Y/Z}-(2{\mathbb{M}}_c)^2}$ in the QCDSR II,  we have to abandon the other solution. In this paper, we retain the mass $M_2$ ($M_{Z^\prime}$) and discard the mass $M_1$ ($M_Z$).

\section{Numerical results and discussions}
We take  the standard values or conventional values  of the vacuum condensates
$\langle\bar{q}q \rangle=-(0.24\pm 0.01\, \rm{GeV})^3$,   $\langle\bar{q}g_s\sigma G q \rangle=m_0^2\langle \bar{q}q \rangle$,
$m_0^2=(0.8 \pm 0.1)\,\rm{GeV}^2$, $\langle \frac{\alpha_s
GG}{\pi}\rangle=(0.33\,\rm{GeV})^4 $    at the typical  energy scale  $\mu=1\, \rm{GeV}$
\cite{SVZ79,Reinders85,ColangeloReview}, and  take the modified minimal subtraction mass of the charm quark $m_{c}(m_c)=(1.275\pm0.025)\,\rm{GeV}$
 from the Particle Data Group \cite{PDG}.
We should evolve the quark condensate, mixed quark condensate and modified minimal subtraction mass  to a special energy scale to warrant the parameters in the QCD spectral densities  having the  same energy scale.    Now let us account for
the energy-scale dependence of  the input parameters at the quark-gluon level,
 \begin{eqnarray}
 \langle\bar{q}q \rangle(\mu)&=&\langle\bar{q}q \rangle({\rm 1GeV})\left[\frac{\alpha_{s}({\rm 1GeV})}{\alpha_{s}(\mu)}\right]^{\frac{12}{33-2n_f}}\, , \nonumber\\
 \langle\bar{q}g_s \sigma G q \rangle(\mu)&=&\langle\bar{q}g_s \sigma G q \rangle({\rm 1GeV})\left[\frac{\alpha_{s}({\rm 1GeV})}{\alpha_{s}(\mu)}\right]^{\frac{2}{33-2n_f}}\, ,\nonumber\\
m_c(\mu)&=&m_c(m_c)\left[\frac{\alpha_{s}(\mu)}{\alpha_{s}(m_c)}\right]^{\frac{12}{33-2n_f}} \, ,\nonumber\\
\alpha_s(\mu)&=&\frac{1}{b_0t}\left[1-\frac{b_1}{b_0^2}\frac{\log t}{t} +\frac{b_1^2(\log^2{t}-\log{t}-1)+b_0b_2}{b_0^4t^2}\right]\, ,
\end{eqnarray}
  where $t=\log \frac{\mu^2}{\Lambda^2}$, $b_0=\frac{33-2n_f}{12\pi}$, $b_1=\frac{153-19n_f}{24\pi^2}$, $b_2=\frac{2857-\frac{5033}{9}n_f+\frac{325}{27}n_f^2}{128\pi^3}$,  with the values $\Lambda=210\,\rm{MeV}$, $292\,\rm{MeV}$  and  $332\,\rm{MeV}$ for the quark  flavors  $n_f=5$, $4$ and $3$, respectively \cite{PDG,Narison-mix}. As we explore the hidden-charm tetraquark states, we choose the flavor $n_f=4$ and search for the best   energy scales   $\mu$.

The Okubo-Zweig-Iizuka supper-allowed decays
\begin{eqnarray}
Z_c&\to&J/\psi\pi\, , \nonumber \\
Z_c^\prime&\to&\psi^\prime\pi\, \nonumber \\
Z_c^{\prime\prime}&\to&\psi^{\prime\prime}\pi\, ,
\end{eqnarray}
are expected to take place easily. The energy gaps maybe have the relations $M_{Z^\prime}-M_{Z}=m_{\psi^\prime}-m_{J/\psi}$ and $M_{Z^{\prime\prime}}-M_{Z^\prime}=m_{\psi^{\prime\prime}}-m_{\psi^\prime}$.
The charmonium masses are $m_{J/\psi}=3.0969\,\rm{GeV}$,   $m_{\psi^\prime}=3.686097\,\rm{GeV}$ and $m_{\psi^{\prime\prime}}=4.039\,\rm{GeV}$ from the Particle Data Group \cite{PDG}, $m_{\psi^\prime}-m_{J/\psi}=0.59\,\rm{GeV}$, $m_{\psi^{\prime\prime}}-m_{J/\psi}=0.94\,\rm{GeV}$, we can  choose the continuum threshold parameters to be  $\sqrt{s_0}=M_{Z}+0.59\,\rm{GeV}$ and $\sqrt{s_0^\prime}=M_{Z}+0.95\,\rm{GeV}$ tentatively and vary the continuum threshold parameters and Borel parameters to satisfy
the following four   criteria:\\
$\bf 1.$ The ground state tetraquark state or single-pole term makes dominant contribution at the hadron  side;\\
$\bf 2.$ The operator product expansion is convergent below the continuum thresholds, and the higher dimensional vacuum condensates make minor contribution;\\
$\bf 3.$  The Borel platforms appear in both the  lineshapes   of the tetraquark masses and pole residues with variations of the Borel parameters;\\
$\bf 4.$ The masses of the tetraquark states satisfy the  energy-scale formula.

 After  trial and error,    we  reach the feasible continuum threshold parameters and  Borel windows, we also acquire the best   energy scales of the  spectral densities at the quark-gluon level and the contributions of the ground state tetraquark states for the QCDSR I, see Table 1.
 In general, for the continuum threshold parameters $s_0$, we can take any  values  satisfy the relation $M_{gr}<\sqrt{s_0}\leq M_{gr}+\Delta$, where the subscript $gr$  denotes  the ground states,
as there exists  an energy gap $\Delta$ between the ground state and the first radial excited state. For the conventional S-wave quark-antiquark mesons, the energy gaps  $\Delta$ vary from $m_{m_{K^*(1410)}}-m_{K^*(892)}=522\,\rm{MeV}$ to $m_{\pi(1300)}-m_{\pi}=1160\,\rm{MeV}$, i.e. $\Delta=522\sim 1160\,\rm{MeV}$ \cite{PDG}.
In the QCD sum rules for the conventional quark-antiquark mesons, we usually choose the values $\sqrt{s_0}=M_{gr}+(0.4\sim0.7)\,\rm{GeV}$ \cite{ColangeloReview}. In Table 1, the continuum threshold parameters $s_0$ satisfy the relation $\sqrt{s_0}=M_{Z_c}+(0.4\sim0.6)\,\rm{GeV}$ or $M_{Z_c}+(0.5\sim0.7)\,\rm{GeV}$, it is reasonable, as the values $\exp\left( -s^0_{max}/T^2_{max}\right)=(1\sim2)\%$, where the subscript $max$ denotes the  maximum values, the contributions of the $Z_c^\prime$ are greatly suppressed if there are any.
In Table 1, we write the continuum threshold parameters as $s_0=21.0\pm 1.0\,\rm{GeV}^2$  rather than as $s_0=(4.58\pm 0.11\,\rm{GeV})^2$ for the $[uc]_{\tilde{A}}[\bar{d}\bar{c}]_A-[uc]_A[\bar{d}\bar{c}]_{\tilde{A}}$ and $[uc]_A[\bar{d}\bar{c}]_A $ tetraquark states to retain the same form as in our previous work \cite{WangEPJC-1601}. In Ref.\cite{WangEPJC-1601}, we study the axialvector $[uc]_A[\bar{d}\bar{c}]_A $ tetraquark state and choose the continuum threshold parameters as $s_0=21.0\pm 1.0\,\rm{GeV}^2$.

 Again we obtain the corresponding parameters for the QCDSR II using trial  and error, see Table 2. In this paper, we employ the energy scale formula $\mu=\sqrt{M^2_{X/Y/Z}-(2{\mathbb{M}}_c)^2}$ with the effective charm quark mass (or constituent charm quark mass) ${\mathbb{M}}_c$ to restrain  the tetraquark masses and the energy scales of the spectral densities  \cite{Wang-tetra-formula}. The energy scale formula can enhance the contributions of the ground state tetraquark states  remarkably at the hadron representation and improve the convergent behaviors of the operator product expansion remarkably at the QCD representation by augmenting the contributions of the lower dimensional vacuum condensates, and is feasible   for  the hidden-charm tetraquark states and hidden-charm pentaquark states \cite{Wang1508-EPJC}.

From Table 1 and Table 2, we can see that the contributions of the single-pole  terms (the ground state tetraquark states)  are about $(40-60)\%$ for the QCDSR I, the contributions of the two-pole  terms (the ground state tetraquark states plus the first radially excited tetraquark states) are about $(70-80)\%$ for the QCDSR II, which satisfy the pole dominance criterion very well.
In the QCDSR II, the contributions of the ground state tetraquark states are about $(30-45)\%$, which are much less than the corresponding ground state tetraquark contributions in the QCDSR I, for the ground state tetraquark masses and pole residues, we prefer the predictions from the QCDSR I.
In numerical calculations, we find  that the contributions of the vacuum condensates of dimension 10 (the largest dimension) are of percent level at the Borel widows for both the QCDSR I and QCDSR II, the minor contributions warrant good convergent behaviors  of  the operator product expansion.

Now let us take into account all the uncertainties of  input parameters, and reach the numerical values of the masses and pole residues of the ground state tetraquark states $Z_c$ and the first radially excited tetraquark states $Z_c^\prime$, which are shown in Table 3 and Table 4. From those Tables, we can see that the ground state tetraquark masses from the QCDSR I and the radially excited  tetraquark  masses from the QCDSR II satisfy the energy scale formula $\mu=\sqrt{M^2_{X/Y/Z}-(2{\mathbb{M}}_c)^2}$, where the updated value of the effective charm quark mass (or constituent charm quark mass) ${\mathbb{M}}_c=1.82\,\rm{GeV}$ is adopted  \cite{WangEPJC-1601}. In Table 4, we also present the central values of the ground state tetraquark masses and pole residues extracted from the QCDSR II at the ideal energy scales which are shown in Table 1. We examine Table 4 and observe  that the ground state tetraquark masses cannot satisfy the energy scale formula, so we will discard those values. This is the shortcoming of the QCDSR II.

\begin{figure}\label{Zc-mass}
\centering
\includegraphics[totalheight=6cm,width=7cm]{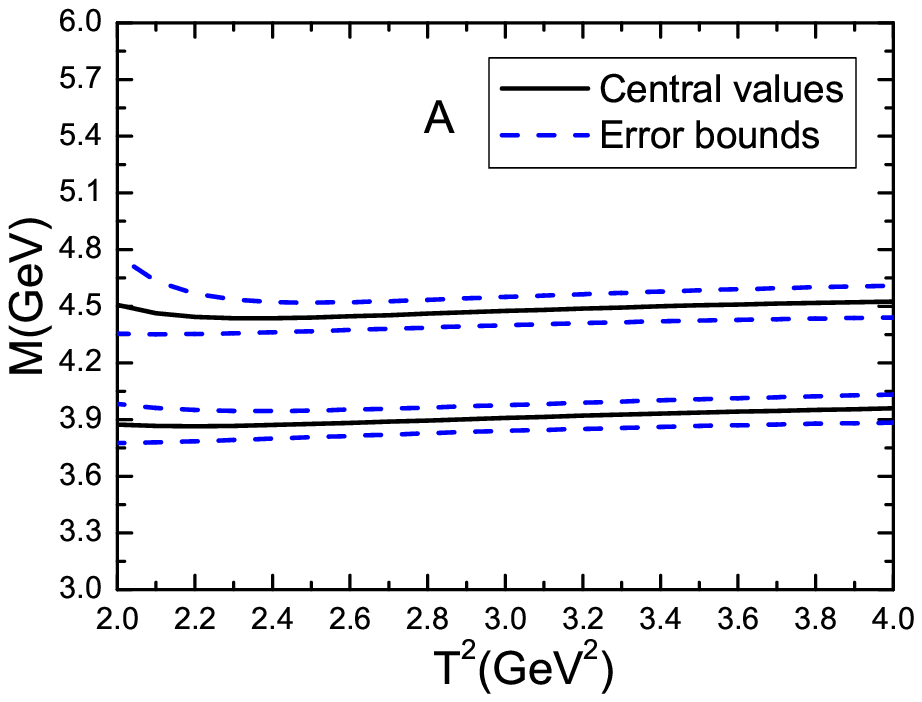}
\includegraphics[totalheight=6cm,width=7cm]{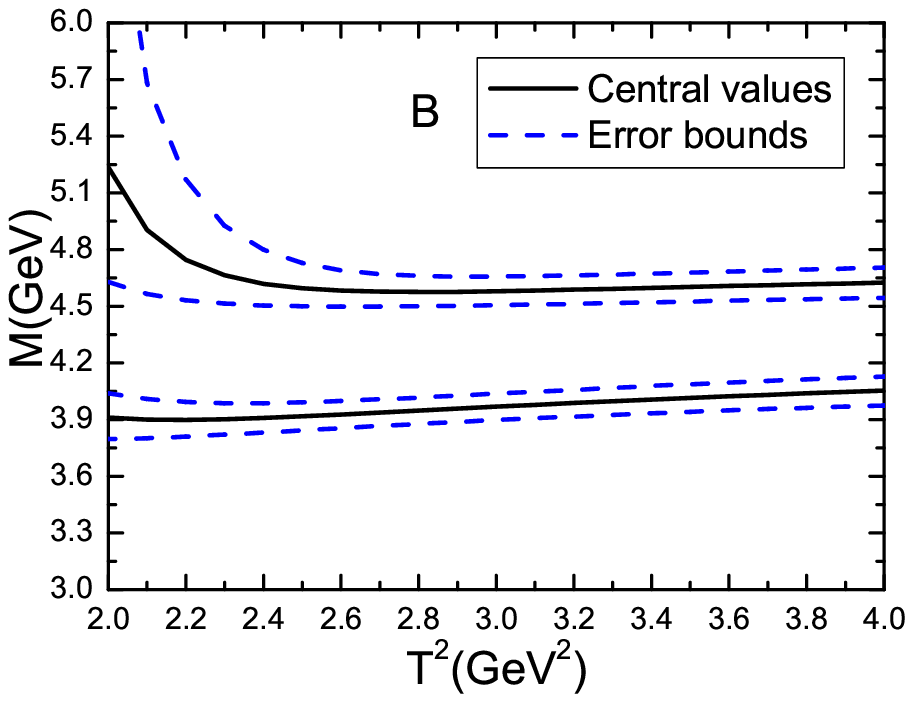}
\includegraphics[totalheight=6cm,width=7cm]{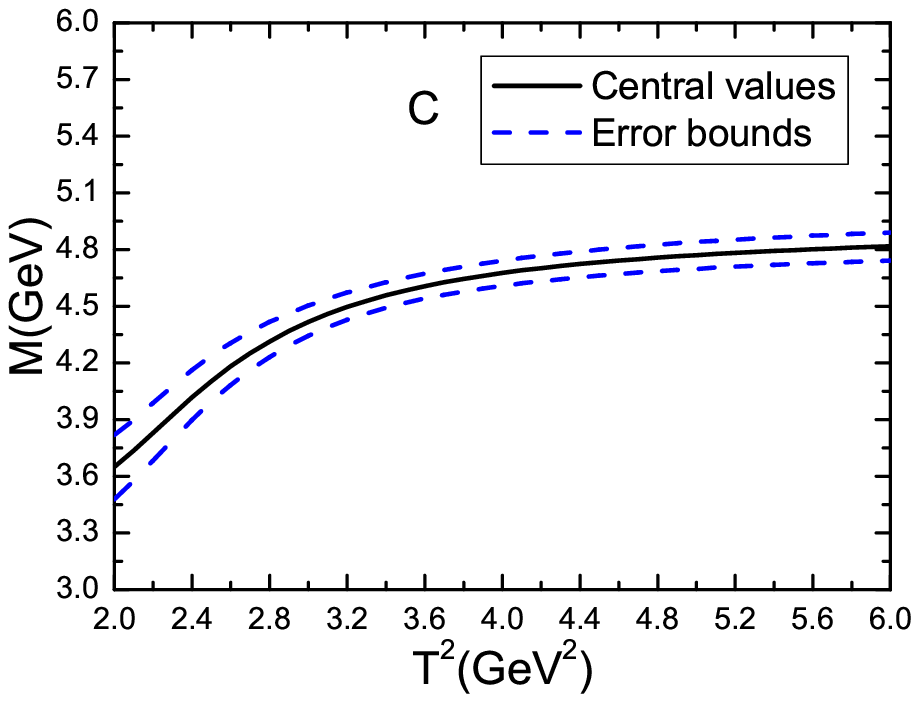}
\includegraphics[totalheight=6cm,width=7cm]{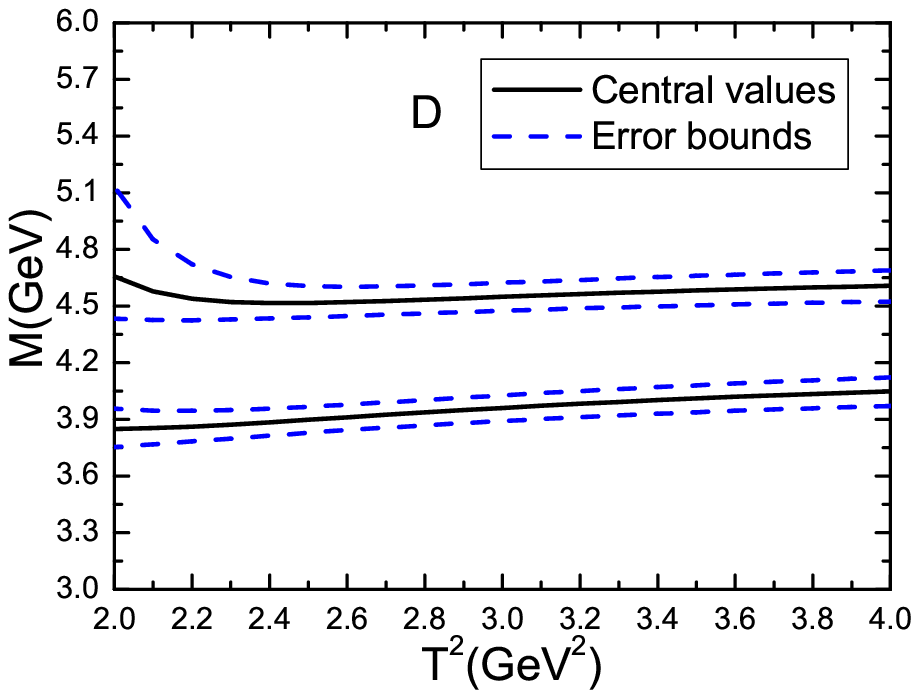}
  \caption{ The masses  with variations of the  Borel parameters $T^2$ for  the axialvector hidden-charm tetraquark states, the $A$, $B$, $C$ and $D$ represent  the $[uc]_S[\bar{d}\bar{c}]_A-[uc]_A[\bar{d}\bar{c}]_S$, $[uc]_{\tilde{A}}[\bar{d}\bar{c}]_A-[uc]_A[\bar{d}\bar{c}]_{\tilde{A}}$, $[uc]_{\tilde{V}}[\bar{d}\bar{c}]_V+[uc]_V[\bar{d}\bar{c}]_{\tilde{V}}$
and $[uc]_A[\bar{d}\bar{c}]_A $ tetraquark states, respectively.  }
\end{figure}

In Fig.\ref{Zc-mass}, we plot the ground state tetraquark masses from the QCDSR I and the first radially excited tetraquark  masses from the QCDSR II in regard to variations of the Borel parameters in much larger regions  than the Borel windows, which are shown in Table 1 and Table 2. From the Fig.\ref{Zc-mass}, we find that there indeed appear very flat platforms in the Borel windows for  the  $[uc]_S[\bar{d}\bar{c}]_A-[uc]_A[\bar{d}\bar{c}]_S$ type, $[uc]_{\tilde{A}}[\bar{d}\bar{c}]_A-[uc]_A[\bar{d}\bar{c}]_{\tilde{A}}$ type and $[uc]_A[\bar{d}\bar{c}]_A $  type axialvector tetraquark states.  For the $[uc]_{\tilde{V}}[\bar{d}\bar{c}]_V+[uc]_V[\bar{d}\bar{c}]_{\tilde{V}}$ type tetraquark state, we only plot the ground state tetraquark mass, as the ground state tetraquark mass is large enough. From the Fig.\ref{Zc-mass}, we also find that the platform in the Borel window is not flat enough, at the region $T^2<3.6\,\rm{GeV}^2$, the mass increases   quickly and monotonously along with the  increase of the value of Borel parameter, the  platform appears  approximately only  at the region $T^2>3.6\,\rm{GeV}^2$.

The predicted mass $M_{Z}=3.90\pm0.08\,\rm{GeV}$ for the ground state tetraquark state $[uc]_S[\bar{d}\bar{c}]_A-[uc]_A[\bar{d}\bar{c}]_S$  exhibits  very good agreement with the experimental value  $M_{Z(3900)}=(3899.0\pm 3.6\pm 4.9)\,\rm{ MeV}$ from the BESIII collaboration \cite{BES3900}, which is in favor of assigning the $Z_c(3900)$ to be the ground state tetraquark state $[uc]_S[\bar{d}\bar{c}]_A-[uc]_A[\bar{d}\bar{c}]_S$  with the quantum numbers  $J^{PC}=1^{+-}$ \cite{WangHuangTao}. In Ref.\cite{WangZhang-Solid}, we study the non-leptonic  decays $Z_c^+(3900)\to J/\psi\pi^+$, $\eta_c\rho^+$, $D^+ \bar{D}^{*0}$, $\bar{D}^0 D^{*+}$ with the three-point  QCD sum rules. In analytical calculations, we take into account both the factorizable  and nonfactorizable Feynman diagrams, match the hadronic representation with the QCD representation according to solid quark-hadron duality, and get  the total decay  width $\Gamma_{Z_c}=54.2\pm29.8\,\rm{MeV}$,
which agrees  with  the experimental value $(46\pm 10\pm 20) \,\rm{MeV}$ very good considering the uncertainties \cite{BES3900}.

 The predicted mass $M_{Z}=4.47\pm0.09\,\rm{GeV}$ for the first radially excited tetraquark state $[uc]_S[\bar{d}\bar{c}]_A-[uc]_A[\bar{d}\bar{c}]_S$  exhibits  very good agreement with the experimental value    $M_{Z(4430)}=(4475\pm7\,{_{-25}^{+15}})\,\rm{ MeV}$ from the   LHCb collaboration \cite{LHCb-1404}, which is in favor of assigning the $Z_c(4430)$ to be the first radially excited tetraquark state  $[uc]_S[\bar{d}\bar{c}]_A-[uc]_A[\bar{d}\bar{c}]_S$  with the quantum numbers  $J^{PC}=1^{+-}$. We can investigate its non-leptonic  decays with the three-point QCD sum rules to make more reasonable assignment.

The predicted mass $M_{Z}=4.01\pm0.09\,\rm{GeV}$ for the ground state tetraquark state $[uc]_{\tilde{A}}[\bar{d}\bar{c}]_A-[uc]_A[\bar{d}\bar{c}]_{\tilde{A}}$
and  $M_{Z}=4.00\pm0.09\,\rm{GeV}$ for the ground state tetraquark state $[uc]_A[\bar{d}\bar{c}]_A$
 both exhibit  very good agreement with the experimental values   $M_{Z(4020/4025)}=(4026.3\pm2.6\pm3.7)\,\rm{MeV}$ \cite{BES1308} and   $(4022.9\pm 0.8\pm 2.7)\,\rm{MeV}$   \cite{BES1309}  from the BESIII collaboration. There are two axialvector tetraquark state candidates with the quantum numbers $J^{PC}=1^{+-}$ for the $Z_c(4020)$. Again the two-body strong decays should be studied to make the assignment more reasonably.

The predicted mass $M_{Z}=4.60\pm0.09\,\rm{GeV}$ for the first radially excited tetraquark state $[uc]_{\tilde{A}}[\bar{d}\bar{c}]_A-[uc]_A[\bar{d}\bar{c}]_{\tilde{A}}$
 and  $M_{Z}=4.58\pm0.09\,\rm{GeV}$ for the first radially excited tetraquark state $[uc]_A[\bar{d}\bar{c}]_A$
 both  exhibit  very good agreement with the experimental value   $M_{Z(4600)}=4600\,\rm{MeV}$ from the LHCb   collaboration \cite{LHCb-Z4600}.
On the other hand, the predicted mass $M_{Z}=4.66\pm0.10\,\rm{GeV}$ for the ground state tetraquark state $[uc]_{\tilde{V}}[\bar{d}\bar{c}]_V+[uc]_V[\bar{d}\bar{c}]_{\tilde{V}}$
is also compatible with the experimental data   $M_{Z(4600)}=4600\,\rm{MeV}$ from the LHCb   collaboration \cite{LHCb-Z4600}. Furthermore, the decay $Z_c(4600)\to J/\psi \pi$
can take place more easily for the ground state tetraquark state, which is in very good agreement  with the observation of the $Z_c(4600)$ in the $J/\psi\pi$ invariant mass spectrum \cite{LHCb-Z4600}.
In summary, there are three axialvector tetraquark state candidates with $J^{PC}=1^{+-}$ for the $Z_c(4600)$, we still need more theoretical and  experimental   works   to identify the $Z_c(4600)$ unambiguously.

In Ref.\cite{Wang-Z4600-V}, we  identify  the $Z_c(4600)$ as  the $[dc]_P[\bar{u}\bar{c}]_A-[dc]_A[\bar{u}\bar{c}]_P$ type vector tetraquark state tentatively according to the predicted mass $M_Z=(4.59 \pm 0.08)\,\rm{GeV}$ from the QCD sum rules \cite{WangY4360Y4660-1803}, and explore
its non-leptonic  decays $Z_c(4600) \to J/\psi \pi$,   $\eta_c \rho$, $J/\psi a_0$,  $\chi_{c0} \rho$, $D^*\bar{D}^*$,  $D\bar{D}$, $D^*\bar{D}$ and $D\bar{D}^*$ with the QCD sum rules by matching the hadronic representation with the QCD representation with solid quark-hadron duality. The large partial decay width $\Gamma(Z_c^-(4600)\to J/\psi \pi^-)=41.4^{+20.5}_{-14.9}\,\rm{MeV}$ exhibits  very good agreement with the observation of the $Z_c(4600)$ in the $J/\psi \pi^-$ invariant mass spectrum.

 In Table \ref{Ground mass}, we also present the diquark spin $S_{uc}$, antidiquark spin  $S_{\bar{d}\bar{c}}$ and  total spin $S$ of the hidden-charm tetraquark states.
We examine the Table and find that the $[uc]_{\tilde{A}}[\bar{d}\bar{c}]_A-[uc]_A[\bar{d}\bar{c}]_{\tilde{A}}$ and $[uc]_A[\bar{d}\bar{c}]_A $ tetraquark states (which have the tetraquark structures $|1^+, 1;  1\rangle-|1, 1^+; 1\rangle$ and $|1, 1;  1\rangle$, respectively) have slightly larger masses than  the $[uc]_S[\bar{d}\bar{c}]_A-[uc]_A[\bar{d}\bar{c}]_S$ tetraquark state (which has the structure $|0, 1;  1\rangle-|1, 0;  1\rangle$). It is reasonable, as the most favored diquark configurations or quark-quark correlations  from the attractive interaction in the color-antitriplet (color-triplet) channel  induced by one-gluon exchange are the scalar diquark (antidiquark) states. In previous works, the scalar, pseudoscalar, vector and axialvector diquarks states have been investigated  with  the QCD sum rules, the scalar and axialvector heavy diquark states  in the color-antitriplet have almost degenerate masses or have almost the same typical quark-quark correlation lengths, the mass gaps between the scalar and axialvector heavy diquark states  are very small or tiny   \cite{WangDiquark}. Furthermore,  it agrees with the predictions of the simple constituent diquark-antidiquark model \cite{Z4430-1405}.

The vector (or P-wave)  diquark states $[uc]_V$ and $[uc]_{\tilde{V}}$ are expected to have larger masses than  the axialvector (or S-wave) diquark states $[uc]_A$ and $[uc]_{\tilde{A}}$, as there exists a relative P-wave between the light quark and heavy quark. In the case of the traditional  $c\bar{u}$ charmed mesons, the energy exciting a P-wave costs about $458\,\rm{MeV}$ from the Particle Data Group \cite{PDG},
\begin{eqnarray}
\frac{5m_{D_2^*}+3m_{D_1}+m_{D^*_0}}{9}-\frac{3m_{D^*}+m_{D}}{4}&=&458\,\rm{MeV}\, .
\end{eqnarray}
If the energy exciting a P-wave in the $qc$ diquark systems also costs about $458\,\rm{MeV}$, the  $[uc]_{\tilde{V}}[\bar{d}\bar{c}]_V+[uc]_V[\bar{d}\bar{c}]_{\tilde{V}}$ tetraquark state has the largest ground state mass, which is even larger than the masses of the first radially  excited states of the  hidden-charm tetraquark states $[uc]_{\tilde{A}}[\bar{d}\bar{c}]_A-[uc]_A[\bar{d}\bar{c}]_{\tilde{A}}$ and  $[uc]_A[\bar{d}\bar{c}]_A $ with the quantum numbers $J^{PC}=1^{+-}$, as exciting two P-waves costs about $0.9\,\rm{GeV}$, which is larger than the energy gap $0.6\,\rm{GeV}$ between the ground state hidden-charm tetraquark state and the first radial excitation of the hidden-charm tetraquark states.

\begin{table}
\begin{center}
\begin{tabular}{|c|c|c|c|c|c|c|c|c|}\hline\hline
 $Z_c$                                                & $T^2 (\rm{GeV}^2)$ & $s_0 $                      & $\mu(\rm{GeV})$   & pole                 \\ \hline
$[uc]_S[\bar{d}\bar{c}]_A-[uc]_A[\bar{d}\bar{c}]_S$   & $2.7-3.1$          & $(4.4\pm0.1\,\rm{GeV})^2$   & $1.4$             & $(40-63)\%$         \\ \hline
$[uc]_{\tilde{A}}[\bar{d}\bar{c}]_A-[uc]_A[\bar{d}\bar{c}]_{\tilde{A}}$   & $3.2-3.6$          & $21.0\pm1.0\,\rm{GeV}^2$    & $1.7$             & $(40-60)\%$         \\ \hline
$[uc]_{\tilde{V}}[\bar{d}\bar{c}]_V+[uc]_V[\bar{d}\bar{c}]_{\tilde{V}}$   & $3.7-4.1$          & $(5.25\pm0.10\,\rm{GeV})^2$ & $2.9$             & $(41-60)\%$         \\ \hline
$[uc]_A[\bar{d}\bar{c}]_A $                           & $3.2-3.6$          & $21.0\pm1.0\,\rm{GeV}^2$    & $1.7$             & $(41-61)\%$         \\ \hline
 \hline
\end{tabular}
\end{center}
\caption{ The Borel parameters, continuum threshold parameters, energy scales of the QCD spectral densities and pole contributions  for the QCDSR I. }
\end{table}

\begin{table}
\begin{center}
\begin{tabular}{|c|c|c|c|c|c|c|c|c|}\hline\hline
 $Z_c+Z^\prime_c$                                     & $T^2 (\rm{GeV}^2)$ & $s_0 $                      & $\mu(\rm{GeV})$   & pole ($Z_c$)                 \\ \hline
$[uc]_S[\bar{d}\bar{c}]_A-[uc]_A[\bar{d}\bar{c}]_S$   & $2.7-3.1$          & $(4.85\pm0.10\,\rm{GeV})^2$ & $2.6$             & $(72-88)\%$ ($(35-52)\%$) \\ \hline
$[uc]_{\tilde{A}}[\bar{d}\bar{c}]_A-[uc]_A[\bar{d}\bar{c}]_{\tilde{A}}$   & $3.2-3.6$          & $(4.95\pm0.10\,\rm{GeV})^2$ & $2.8$             & $(64-80)\%$ ($(30-44)\%$)\\ \hline
$[uc]_A[\bar{d}\bar{c}]_A $                           & $3.2-3.6$          & $(4.95\pm0.10\,\rm{GeV})^2$ & $2.8$             & $(64-81)\%$ ($(29-43)\%$)   \\ \hline
 \hline
\end{tabular}
\end{center}
\caption{ The Borel parameters, continuum threshold parameters, energy scales of the QCD spectral densities  and pole contributions  for the QCDSR II. }
\end{table}

\begin{table}
\begin{center}
\begin{tabular}{|c|c|c|c|c|c|c|c|c|}\hline\hline
 $Z_c$                                                &$|S_{uc}, S_{\bar{d}\bar{c}}; S\rangle$    &$M_Z (\rm{GeV})$   &$\lambda_Z (\rm{GeV}^5)$    \\ \hline

$[uc]_S[\bar{d}\bar{c}]_A-[uc]_A[\bar{d}\bar{c}]_S$   &$|0, 1;  1\rangle-|1, 0;  1\rangle$        &$3.90\pm0.08$      &$(2.09\pm0.33)\times 10^{-2}$    \\ \hline

$[uc]_{\tilde{A}}[\bar{d}\bar{c}]_A-[uc]_A[\bar{d}\bar{c}]_{\tilde{A}}$   &$|1^+, 1;  1\rangle-|1, 1^+; 1\rangle$     &$4.01\pm0.09$      &$(5.96\pm0.94)\times 10^{-2}$  \\ \hline

$[uc]_{\tilde{V}}[\bar{d}\bar{c}]_V+[uc]_V[\bar{d}\bar{c}]_{\tilde{V}}$   &$|1^-, 1;  1\rangle+|1, 1^-; 1\rangle$     &$4.66\pm0.10$      &$(1.18\pm0.22)\times 10^{-1}$   \\ \hline

$[uc]_A[\bar{d}\bar{c}]_A $                           &$|1, 1;  1\rangle$                         &$4.00\pm0.09$      &$(2.91\pm0.46)\times 10^{-2}$  \\ \hline \hline
\end{tabular}
\end{center}
\caption{ The masses and pole residues of the ground state tetraquark states $Z_c$  from the QCDSR I, where the superscripts $\pm$ represent  the positive parity and negative parity constituents of the tensor diquark states, respectively. }\label{Ground mass}
\end{table}

\begin{table}
\begin{center}
\begin{tabular}{|c|c|c|c|c|c|c|c|c|}\hline\hline
 $Z_c+Z_c^\prime$                                    &$M_Z (\rm{GeV})$  &$\lambda_Z (\rm{GeV}^5)$  &$M_{Z^\prime}(\rm{GeV})$ &$\lambda_{Z^\prime}(\rm{GeV}^5)$                \\ \hline
$[uc]_S[\bar{d}\bar{c}]_A-[uc]_A[\bar{d}\bar{c}]_S$  &$3.81 $           &$1.77\times10^{-2}$       &$4.47\pm0.09$            &$(6.02\pm0.80)\times10^{-2}$ \\ \hline
$[uc]_{\tilde{A}}[\bar{d}\bar{c}]_A-[uc]_A[\bar{d}\bar{c}]_{\tilde{A}}$  &$3.78$            &$3.94\times10^{-2}$       &$4.60\pm0.09$            &$(1.35\pm0.18)\times10^{-1}$  \\ \hline
$[uc]_A[\bar{d}\bar{c}]_A $                          &$3.73$            &$1.76\times10^{-2}$       &$4.58\pm0.09$            &$(6.55\pm0.85)\times10^{-2}$ \\ \hline
 \hline
\end{tabular}
\end{center}
\caption{ The masses and pole residues of the ground state tetraquark states $Z_c$ and the first radially excited tetraquark states $Z_c^\prime$ from the QCDSR II. }
\end{table}

\section{Conclusion}
In this paper, we investigate  the ground states and the first radially excited states of the $[uc]_S[\bar{d}\bar{c}]_A-[uc]_A[\bar{d}\bar{c}]_S$ type,
$[uc]_{\tilde{A}}[\bar{d}\bar{c}]_A-[uc]_A[\bar{d}\bar{c}]_{\tilde{A}}$ type and $[uc]_A[\bar{d}\bar{c}]_A $ type tetraquark states  and  the ground state $[uc]_{\tilde{V}}[\bar{d}\bar{c}]_V+[uc]_V[\bar{d}\bar{c}]_{\tilde{V}}$ type tetraquark state with the quantum numbers $J^{PC}=1^{+-}$ via the QCD sum rules in a systematic way.
The predicted tetraquark masses are in favor of assigning the $Z_c(3900)$ and $Z_c(4430)$ as the ground state and the first radially excited state of the $[uc]_S[\bar{d}\bar{c}]_A-[uc]_A[\bar{d}\bar{c}]_S$ type axialvector tetraquark states respectively;
assigning the $Z_c(4020)$ as the ground state $[uc]_{\tilde{A}}[\bar{d}\bar{c}]_A-[uc]_A[\bar{d}\bar{c}]_{\tilde{A}}$ type  axialvector tetraquark state
or $[uc]_A[\bar{d}\bar{c}]_A$ type axialvector tetraquark state; assigning the $Z_c(4600)$ as  the first radially excited $[uc]_{\tilde{A}}[\bar{d}\bar{c}]_A-[uc]_A[\bar{d}\bar{c}]_{\tilde{A}}$ type axialvector tetraquark state or $[uc]_A[\bar{d}\bar{c}]_A$ type axialvector tetraquark state, or the ground state $[uc]_{\tilde{V}}[\bar{d}\bar{c}]_V+[uc]_V[\bar{d}\bar{c}]_{\tilde{V}}$ type axialvector tetraquark state. We still need more experimental and theoretical works to identify the $Z_c(4600)$ unambiguously.

\section*{Acknowledgements}
This  work is supported by National Natural Science Foundation, Grant Number  11775079.

\end{document}